\documentclass[twocolumn]{aastex63}
\usepackage{graphicx}

\shorttitle{LcTools}
\shortauthors{Schmitt, Hartman, and Kipping}
\graphicspath{ {./images/} }

\begin{document}


\title{LcTools: A Windows-Based Software System for Finding and Recording Signals in Lightcurves from NASA Space Missions}

\author{Allan R. Schmitt}
\affiliation{Citizen Scientist, 616 W. 53rd. St., Apt. 101, Minneapolis, MN 55419, USA, \href{mailto:aschmitt@comcast.net}{aschmitt@comcast.net}}

\author{Joel D. Hartman}
\affiliation{Department of Astrophysical Sciences, Princeton University, 4 Ivy Ln, Princeton, NJ 08544, USA, \href{mailto:jhartman@astro.princeton.edu}{jhartman@astro.princeton.edu}}

\author{David M. Kipping}
\affiliation{Department of Astronomy, Columbia University, 550 W 120th Street, New York, NY 10027, USA, \href{mailto:dkipping@astro.columbia.edu}{dkipping@astro.columbia.edu}}



\begin{abstract}

Since 2009, the Kepler, K2, and TESS missions have produced a vast number of lightcurves for public use. To assist citizen scientists in processing those lightcurves, the LcTools software system was developed. The system provides a set of tools to efficiently search for signals of interest in large sets of lightcurves using automated and manual (visual) techniques. At the heart of the system is a multipurpose lightcurve viewer and signal processor with advanced navigation and display capabilities to facilitate the search for signals. Other applications in the system are available for building lightcurve files in bulk, finding periodic signals automatically, and generating signal reports. This paper describes each application in the system and the methods by which the software can be used to detect and record signals. The software is free and can be obtained from the lead author by request at \href{mailto:aschmitt@comcast.net}{aschmitt@comcast.net}.
\end{abstract}



\section{Introduction} \label{sec:intro} 

LcTools is a Windows based software system for finding and recording signals in the lightcurves for supported projects and associated High Level Science Products (HLSPs) at MAST. Supported projects include TESS, K2, and Kepler. Supported HLSPs include TASOC, K2SFF, and EVEREST. 

A signal may be recorded for any type of astronomical event, artifact, or anomaly detected in a lightcurve whether periodic or non-periodic. Examples of transit based signals include planets, eclipsing binaries, moons, rings, trojans, and comets.

The system consists of four major applications \textbf{--} LcViewer, LcSignalFinder, LcGenerator, and LcReporter. \textbf{LcViewer} is a multipurpose lightcurve viewer and signal processor enabling a user to 1) generate, edit, and detrend lightcurves, 2) detect, record, measure, track, locate, query, and display signals, 3) import and display project based signals such as TOIs and KOIs, 4) record TTVs, 5) phase fold periodic signals, 6) measure time and flux intervals, and 7) query stellar properties. \textbf{LcSignalFinder} automatically detects and records periodic signals found in large sets of lightcurves. \textbf{LcGenerator} builds lightcurve files in bulk for subsequent use with LcViewer and LcSignalFinder. \textbf{LcReporter} creates an Excel report for the signals recorded by LcViewer.

The software system was primarily developed by A. S. over nine years starting in 2011. J. H. contributed the BLS (Box-Fitting Least Squares) \citep{Kovacs} module derived from VARTOOLS \citep{Hartman} for automatically finding periodic signals in lightcurves. D. K. contributed the underlying algorithms and design for detrending lightcurves and phase folding signals.

Although LcTools may be used by anyone, from novices to professionals, it is primarily intended for advanced citizen scientists, students, and universities. The LcTools community currently consists of 77 registered users worldwide. A recent survey of the publications posted on arXiv\footnote{\url{https://arxiv.org/list/astro-ph.EP/recent}} found 20 papers that either cited or acknowledged use of the product for investigating various astronomical phenomena \citep{2019MNRAS.488.4520C,2019MNRAS.488.4465G,2019MNRAS.488.2455R,2019arXiv190909094E,2019MNRAS.485.2681R,2019MNRAS.483.3579A,2019MNRAS.483.1934B,2019AJ....157...17L,2018A&A...620A.189O,2018AJ....156..245R,2018MNRAS.478.5135B,2018AJ....155..107M,2018RNAAS...2...28L,2018MNRAS.474.1453R,2018ApJ...854..109Z,2018AJ....155...57C,2017MNRAS.467.2160R,2016AJ....151..159S,2016ApJ...816...69A,2015ApJ...813...14K}.

This paper is organized as follows. Section \ref{sec:projects} lists the projects and HLSPs supported in LcTools. Section \ref{sec:assets} describes the data assets that can be used with the system. Section \ref{sec:generator} covers LcGenerator while section \ref{sec:finder} describes LcSignalFinder. Sections \ref{sec:viewer} and \ref{sec:reporter} cover LcViewer and LcReporter respectively. Section \ref{sec:summary} provides a summary and concluding remarks.

See Appendix \ref{sec:rtreqs} for a list of hardware and software requirements for running LcTools.


\section{Supported Projects and HLSPs} \label{sec:projects}

LcTools currently supports three projects and three HLSPs. Projects include:
\begin{itemize}
\item TESS \citep{Ricker} \textbf{--} 2-minute cadence lightcurves for 200,000 stars from the Candidate Target List (CTL).
\item K2 \citep{Howell} \textbf{--} Long and short cadence lightcurves for 330,000 stars observed from 2014-2018 spanning 19 campaigns.
\item Kepler \citep{Koch} \textbf{--} Long and short cadence lightcurves for 200,000 stars observed from 2009-2013 spanning 17 quarters.
\end{itemize}

HLSPs include:
\begin{itemize}
\item TASOC \citep{Handberg} \textbf{--} Long cadence lightcurves produced from TESS Full-Frame Images (FFIs).
\item K2SFF \citep{Vanderburg} \textbf{--} Corrected long cadence K2 lightcurves spanning 19 campaigns.
\item EVEREST \citep{Luger} \textbf{--} Corrected long and short cadence K2 lightcurves spanning 19 campaigns.
\end{itemize}

In addition, TESS FFI lightcurve files for sector 1 are available from Oelkers \citep{Oelkers_2018} as obtained from the TESS Full Frame Image Portal\footnote{\url{https://filtergraph.com/tess_ffi/sector-01}}.

\section{Data Assets} \label{sec:assets}

Five types of data assets can be used with LcTools \textbf{--} lightcurve files, star list files, signal libraries, TTV libraries, and stellar properties. A description of each type is provided below.

\subsection{Lightcurve Files}

A lightcurve file contains time series information for a star for use with LcSignalFinder and LcViewer. Lightcurve files can be built using LcGenerator or LcViewer from the data archived at MAST\footnote{\url{https://archive.stsci.edu/}} (the Mikulski Archive for Space Telescopes). Lightcurve files can also be downloaded directly from the LcTools website\footnote{\url{https://sites.google.com/a/lctools.net/lctools/data-sources}}. Lightcurve directories on the LcTools website are available by sector for TESS, by campaign for K2SFF, and by batch for Kepler.

\subsection{Star List Files}

A star list file contains a list of stars observed in a specific time period or batch. Time periods are project specific \textbf{--} sectors for TESS and campaigns for K2. A star list file can be downloaded from the LcTools website and fed to LcGenerator for building a custom lightcurve directory. Multi-period star lists can also be downloaded from the website. 

\subsection{Signal Libraries} \label{subsec:siglibs}

A signal library is a set of signals that can be imported by LcSignalFinder and LcViewer when a lightcurve file is loaded. Signal libraries may be project, public, or private.

A project signal library contains planet candidate signals defined by the project. Libraries include:
\begin{itemize}
\item TOIs \textbf{--} TESS Objects of Interest obtained from ExoFOP-TESS\footnote{\url{https://exofop.ipac.caltech.edu/tess/}}.
\item CTOIs \textbf{--} Community TESS Objects of Interest obtained from ExoFOP-TESS.
\item K2OIs \textbf{--} K2 Objects of Interest obtained from NEA\footnote{\url{https://exoplanetarchive.ipac.caltech.edu/}} (the NASA Exoplanet Archive).
\item KOIs \textbf{--} Kepler Objects of Interest obtained from NEA.
\item TCEs \textbf{--} Threshold Crossing Events obtained from MAST for TESS and from NEA for Kepler.
\end{itemize}

A public signal library contains signals that were created with LcViewer for use by a group of individuals as part of a team collaboration. The signals are located in a shared Google Drive folder.

A private signal library contains signals that were created with LcViewer for personal use only. The signals are normally located in the ``Signals" subdirectory of a lightcurve directory.

\subsection{TTV Libraries} \label{subsec:ttvlibs}

A TTV library contains information for aligning periodic signals in a lightcurve that have been shifted due to transit timing variations. A library can be imported by LcSignalFinder and LcViewer when a lightcurve file is loaded. TTV libraries may be public or private. Use of TTV libraries is optional.

Currently, there are two public TTV libraries available in shared Google Drive folders. Both are for the Kepler project. They include:
\begin{itemize}
\item TTVs\_Kepler\_Mazeh \textbf{--} A TTV library derived from the Holczer and Mazeh catalog \citep{Holczer}.
\item TTVs\_Kepler\_Schmitt \textbf{--} A supplemental TTV library manually created with LcViewer by the lead author for use with TTVs\_Kepler\_Mazeh. 
\end{itemize}

\subsection{Stellar Properties} \label{subsec:stellarprops}

Stellar properties may be imported from MAST or NEA by LcSignalFinder and LcViewer when a lightcurve file is loaded. Properties include the star ID, stellar magnitude, temperature, mass, radius, and distance to the star.

\section{\large{\large{L\lowercase{c}G\lowercase{enerator}}}} \label{sec:generator}

LcGenerator builds lightcurve files in bulk for use with LcSignalFinder and LcViewer. For example, LcGenerator can be used to build lightcurve files for all the CTL stars in a TESS sector typically consisting of 20,000 stars.

The main application window for LcGenerator is shown in Figure \ref{fig:LcGenerator}. The window can be moved anywhere on the screen. The position will be remembered the next time the application is started. The window can also be minimized so that the application runs in the background out of the way.

\subsection{Setting Up a Job} \label{subsec:gensetup}

Main job settings in the application window include:

\begin{figure*}[htb!]
\includegraphics[scale=0.75]{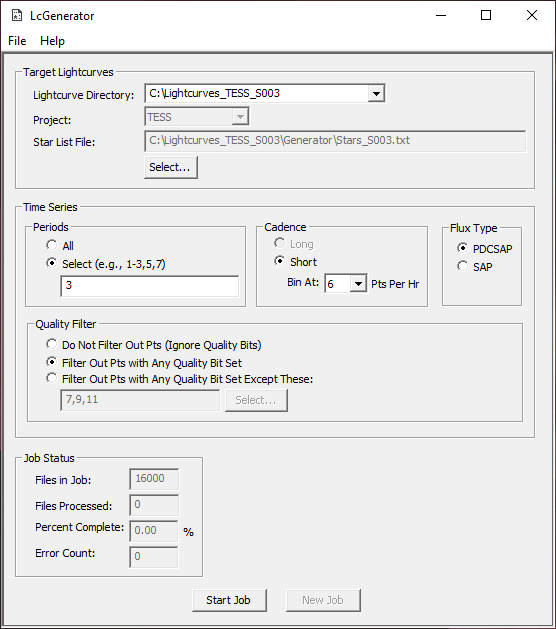}
\centering
\caption{The LcGenerator Window.}
\label{fig:LcGenerator}
\end{figure*}

\begin{itemize}
\item The target lightcurve directory.
\item The LcTools project name. Available project names include TESS, TESS\_TASOC, K2, K2\_EVEREST, K2\_SFF, and Kepler.
\item The star list file for lightcurve files to build. Typically, a star list file is downloaded from the LcTools website.
\item The time series periods to include in the build. Periods are project specific \textbf{--} sectors for TESS, campaigns for K2, and quarters for Kepler.
\item The desired cadence type \textbf{--} long or short. For short cadence data, the bin size must also be specified in data points per hour (PPH). Available bin sizes include 2, 3, 4, 5, 6, 10, 12, 15, 20, and 30 PPH for TESS related projects and 2, 3, 4, 5, 6, 10, 12, 15, 20, 30, and 60 PPH for K2 and Kepler related projects.
\item The flux type \textbf{--} PDCSAP for the detrended/corrected data and SAP for the raw/uncorrected data.
\item The quality filter for removing low quality data points from the generated lightcurves. If radio button 1 is selected, no data points will be removed. If button 2 is selected, all data points having flagged quality bits will be removed. If button 3 is selected, all data points having flagged quality bits will be removed except a selected list of quality bits to ignore.
\end{itemize}

\subsection{Executing a Job} \label{subsec:genexec}

To begin a job, the ``Start Job" button must be clicked. At any point thereafter, the job can be paused and resumed by clicking the associated on-screen buttons (not shown). Progress of the job can be monitored via the Job Status box located in the bottom left corner of the window.

\subsection{The Build Process for a Lightcurve File} \label{subsec:genbldprocess}

For each lightcurve file to build, LcGenerator performs the following tasks subject to the job settings: 1) Downloads the applicable time series files from MAST, 2) converts the time series files from FITS to text, 3) filters out low quality data points, 4) for short cadence data, bins the data points to the specified data rate, 5) normalizes the flux values to a mean value of 1.0, 6) merges the normalized time series data together, and 7) writes the resulting file into the lightcurve directory.

\subsection{Supporting Documentation}

A user guide for LcGenerator can be accessed from the menu bar. The document can also be accessed from the LcTools installation directory.

\section{\large{\large{L\lowercase{c}S\lowercase{ignal}F\lowercase{inder}}}} \label{sec:finder}

LcSignalFinder uses the BLS algorithm to detect and record periodic signals found in a large set of lightcurve files for subsequent use with LcViewer. For example, LcSignalFinder can be used to detect and record all the periodic signals found in the CTL lightcurve files for a TESS sector typically consisting of 20,000 files. Detection of both periodic dips and periodic peaks is supported.

The main application window for LcSignalFinder is shown in Figure \ref{fig:LcSignalFinder}. The window can be moved anywhere on the screen. The position will be remembered the next time the application is started. The window can also be minimized so that the application runs in the background out of the way.

\subsection{Setting Up a Job} \label{subsec:findersetup}

\subsubsection{Main Job Settings} \label{subsubsec:findermainsetup}

Main job settings in the application window include:

\begin{figure*}[htb!]
\includegraphics[scale=0.70]{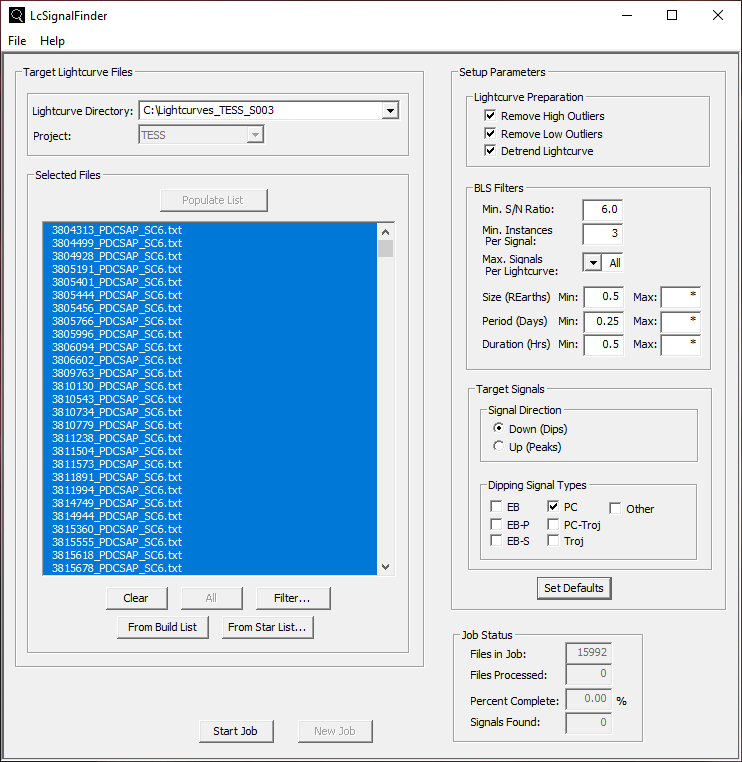}
\centering
\caption{The LcSignalFinder Window}
\label{fig:LcSignalFinder}
\end{figure*}

\begin{itemize}
\item The lightcurve directory in which to search for periodic signals. Typically, this directory is either downloaded from the LcTools website or built with LcGenerator. 
\item The list of lightcurve files to process in the directory. By default, all files are selected. Individual files can be selected or deselected with the mouse. 
  
The ``Filter" button selects files on the basis of a specified search pattern. For example, ``*PDCSAP*" selects all files with ``PDCSAP" in the filename.

The ``From Build List" button selects all files that were last built with LcGenerator. 

The ``From Star List" button selects all files whose leading star IDs match the star IDs found in a specified star list file.
\item The lightcurve preparation options. Checkboxes are available for removing high data point outliers, removing low data point outliers, and for detrending the lightcurves. Normally, all three boxes are checked.
\item BLS filters for signals to find and return. Filters include 1) min. signal-to-noise ratio, 2) min. number of instances per periodic signal, 3) max. number of signals per lightcurve, 4) min. and max. signal size in REarth units, 5) min. and max. period in days, and 6) min. and max. signal duration in hours.
\item The target signal direction \textbf{--} Down for dips and Up for peaks.
\item The type(s) of dipping signals to find. Seven types are available:

\begin{enumerate}
\item EB \textbf{--} Eclipsing binary signals.
\item EB-P \textbf{--} Primary eclipse signals for EBs.
\item EB-S \textbf{--} Secondary eclipse signals for EBs.
\item PC \textbf{--} Planet candidate signals.
\item PC-Troj \textbf{--} Host planet signals for trojans.
\item Troj \textbf{--} Trojan signals.
\item Other \textbf{--} All other periodic signals.
\end{enumerate}

\end{itemize}

\subsubsection{Signal Libraries} \label{subsubsec:findersiglibssetup}

\begin{figure}[ht]
\includegraphics[scale=0.75]{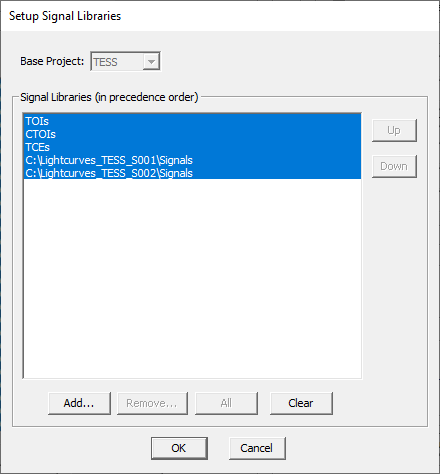}
\caption{The ``Setup Signal Libraries" Dialog Box}
\label{fig:SetupSigLibs}
\end{figure}

Signal libraries for the job can be set up using the dialog box shown in Figure \ref{fig:SetupSigLibs}.

At minimum, the library list will contain all available project libraries for the base project \textbf{--} TOIs, CTOIs, and TCEs for TESS, K2OIs for K2, and KOIs and TCEs for Kepler. Optionally, there may be one or more public and private signal libraries in the list. Normally all available signal libraries are selected.

In the example shown, there are five libraries in the list all of which are selected for use. The first three are project libraries. The last two are private libraries. Using the mouse, the user can select and deselect the desired libraries as needed.

The list is maintained in precedence order with the highest priority library at the top and the lowest priority library at the bottom. Items in the list can be reordered using the ``Up" and ``Down" buttons.

Public and private libraries can be added to the list or removed from the list via the associated buttons. Project libraries cannot be added or removed.

\subsubsection{TTV Libraries} \label{subsubsec:finderttvlibssetup}

TTV libraries for the job can be set up using a dialog box similar to the one for signal libraries. Public and private libraries can selected, deselected, reordered, added, and removed from the list as needed.

\subsubsection{Stellar Properties} \label{subsubsec:finderstellarprops}

\begin{figure}[ht]
\includegraphics[scale=0.75]{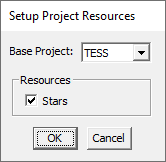}
\centering
\caption{The ``Setup Project Resources" Dialog Box}
\label{fig:SetupProjResDlg}
\end{figure}

Stellar properties for the job can be enabled or disabled using the dialog box shown in Figure \ref{fig:SetupProjResDlg}. Normally, stellar properties are enabled.

\subsection{Executing a Job} \label{subsec:finderexec}

To begin a job, the ``Start Job" button must be clicked. At any point thereafter, the job can be paused and resumed by clicking the associated on-screen buttons (not shown). Progress of the job can be monitored via the Job Status box located in the bottom right corner of the window.

\subsection{The Signal Detection Process for a Lightcurve File} \label{subsubsec:finderprocess}

For each selected lightcurve file, LcSignalFinder performs the following tasks subject to the job settings:

\begin{enumerate}
\item Reads the lightcurve file into memory.
\item Imports the stellar properties for the host star from MAST or NEA.
\item Imports the signals for the lightcurve from the selected signal libraries in precedence order and then instantiates them in the lightcurve.
\item Imports the TTV records for the signals from the selected TTV libraries in precedence order and then aligns the instances of the signals in the lightcurve based on the offset information.
\item Removes the data points for all instantiated signals in the lightcurve so that BLS will not find and return known signals.
\item Removes the worst 1\% of high outliers and the worst 0.05\% of low outliers from the lightcurve.
\item Detrends the lightcurve using a moderately aggressive moving median curve fit.
\item If the target signal direction is Up, inverts the data points in the lightcurve so that peaks become dips enabling BLS to detect them.
\item Executes BLS repeatedly to find all periodic signals in the lightcurve. Processing stops when the maximum number of signals is reached per the job settings or no more signals are found.
\item Analyzes the signals found. If the target signal direction is Down, classifies each signal as follows:

\begin{itemize}
\item If the signal size is greater than 30 REarths or the signal depth is greater than 50,000 ppm if the stellar radius is unknown, the signal is classified as an EB. Otherwise it is classified as a PC.
\item If two PC signals have nearly identical periods and have reference epochs that differ by approximately 1/6th of a period (corresponding to the L4 and L5 Lagrange points), the deeper PC signal is re-classified as a PC-Troj and the shallower PC signal is re-classified as a Troj.
\item If two EB or PC signals have nearly identical periods, the deeper signal is re-classified as an EB-P (primary eclipse) and the shallower signal is re-classified as an EB-S (secondary eclipse).
\item If the signal duration is greater than 0.2 times the signal period, the signal is classified as Other. Such signals are not likely to be planets or EBs.
\end{itemize}

The classification method described above only serves as an initial approximation based on preliminary information. Results should not be regarded as absolute. Vetting of the signals will be required before a more accurate classification can be determined.
\item Records all periodic signals that match the job settings in the LcSignalFinder window. 
\end{enumerate}

\subsection{Vetting the Signals} \label{subsec:findervet}

Vetting of the periodic signals recorded by LcSignalFinder is handled in LcViewer as described in section \ref{subsec:vettingsigs}. Both LcSignalFinder and LcViewer can be run concurrently to expedite the processing of signals in a lightcurve directory.

\subsection{Supporting Documentation}

A user guide for LcSignalFinder can be accessed from the menu bar. The document can also be accessed from the LcTools installation directory.

\section{\large{L\lowercase{c}V\lowercase{iewer}}} \label{sec:viewer}

LcViewer is a multipurpose graphics application for finding and recording signals of interest in lightcurves. Via the application, a user is able to 1) build, view, edit, and detrend lightcurves, 2) detect, record, measure, track, locate, query, display, and phase fold signals, 3) record TTVs, 4) show background flux, 5) measure time and flux intervals, and 6) query stellar properties.

See Figure \ref{fig:MenuBar} for available menu bar commands and Figure \ref{fig:Hotkeys} for available hot keys and hot buttons. Although many operations in LcViewer may be performed using either technique, for brevity only hot keys and hot buttons will be mentioned in this document.

In the discussion that follows, M1 refers to the left mouse button and M2 to the right mouse button.

\subsection{Setting Up the Environment}

Signal libraries, TTV libraries, and stellar properties for LcViewer can be set up in a manner similar to LcSignalFinder. See sections \ref{subsubsec:findersiglibssetup}, \ref{subsubsec:finderttvlibssetup}, and \ref{subsubsec:finderstellarprops} respectively.

\subsection{Opening Lightcurve Files}

Lightcurve files can be opened in two ways \textbf{--} via a work group for opening a large set of files sequentially and through a dialog box for opening files individually.


\subsubsection{Opening Lightcurve Files Through a Work Group} \label{subsec:workgroup}

A work group is a set of lightcurve files to open sequentially at the click of a button or press of a key for rapidly iterating though a large list of files. For example, a work group could be set up for viewing all the CTL lightcurve files for a TESS sector typically consisting of 20,000 files.

\begin{figure}[ht]
\includegraphics[scale=0.75]{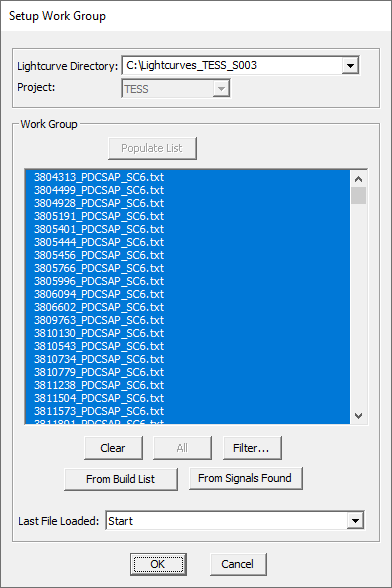}
\centering
\caption{The ``Setup Work Group" Dialog Box}
\label{fig:SetupWorkGroupDlg}
\end{figure}

A work group can be set up by clicking the ``Setup" button located in the lower left corner of the LcViewer window (see Figure \ref{fig:LcViewer}) or by pressing the Shift+w key. The ``Setup Work Group" dialog box will then be opened as shown in Figure \ref{fig:SetupWorkGroupDlg}. 

By default, all the files in a specified lightcurve directory are selected for use. Individual items in the list can be selected or deselected using the mouse.

Clicking the ``Filter" button selects files on the basis of a specified search pattern. For example, ``*PDCSAP*" selects all files with ``PDCSAP" in the filename.

Clicking the ``From Build List" button selects all files that were last built with LcGenerator. 

Clicking the ``From Signals Found" button selects all files in which periodic signals were found by LcSignalFinder.  This step is required as part of the signal vetting process as described in section \ref{subsec:vettingsigs}. Files with signals that have already been dispositioned (created or deleted) as part of the vetting process are excluded from the selected list.

Once a work group has been set up, the files can be opened sequentially (forward or backward) by clicking the ``Next" or ``Prev" button located in the lower left corner of the LcViewer window (see Figure \ref{fig:LcViewer}) or by pressing the equivalent Shift+n or Shift+p key.

The loading process for a lightcurve file is described in section \ref{subsubsec:lcloadprocess}.


\subsubsection{Opening Lightcurve Files Through a Dialog Box}

\begin{figure}[ht]
\includegraphics[scale=0.75]{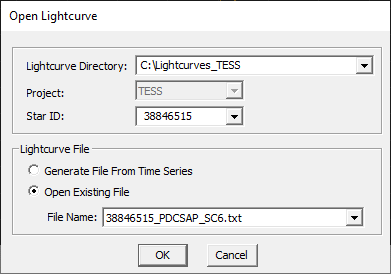}
\centering
\caption{The ``Open Lightcurve" Dialog Box}
\label{fig:OpenLcDlg}
\end{figure}

A lightcurve file can be opened individually by pressing the Shift+f key or by selecting the equivalent command from the menu bar. The ``Open Lightcurve" dialog box will then be displayed as shown in Figure \ref{fig:OpenLcDlg}. Upon selecting the target lightcurve directory and host star ID, the method for opening the file must be selected. Two options are available \textbf{--} generate a new file in the directory or open an existing file from the directory.

\begin{figure}[ht]
\includegraphics[scale=0.7]{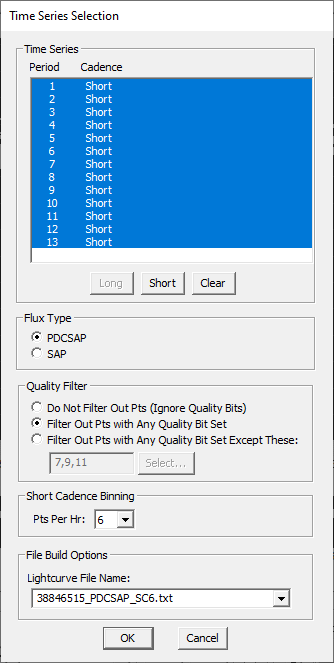}
\centering
\caption{The ``Time Series Selection" Dialog Box}
\label{fig:TimeSeriesSelectionDlg}
\end{figure}

If the first option is selected, the ``Time Series Selection" dialog box will be opened for specifying the build options as shown in Figure \ref{fig:TimeSeriesSelectionDlg}. The build options and build process for a lightcurve file are similar to those described for LcGenerator and so will not be repeated here. See sections \ref{subsec:gensetup} and \ref{subsec:genbldprocess} respectively.

\newpage
\subsubsection{The Load Process for a Lightcurve File} \label{subsubsec:lcloadprocess}
Once a lightcurve file is ready to be opened, LcViewer performs the following tasks subject to the environment settings:

\begin{enumerate}
\item Reads the file into memory.
\item Imports the stellar properties for the host star from MAST or NEA.
\item Imports the signals for the lightcurve from the selected signal libraries in precedence order and then instantiates them in the lightcurve.
\item Imports the TTV records for the signals from the selected TTV libraries in precedence order and then aligns the instances of the signals in the lightcurve based on the offset information.
\item Displays the lightcurve in the main LcViewer window.
\end{enumerate}


\newpage
\subsection{The LcViewer Window} \label{subsec:viewerwndo}

The main LcViewer window is shown in Figure \ref{fig:LcViewer}. The window can be resized and repositioned anywhere on the screen. The size and position will be remembered the next time the application is started. The window can be maximized to fill the screen or minimized so that it runs in the background out of the way.

Major components in the window include:

\begin{figure*}[htb!]
\includegraphics[scale=0.9]{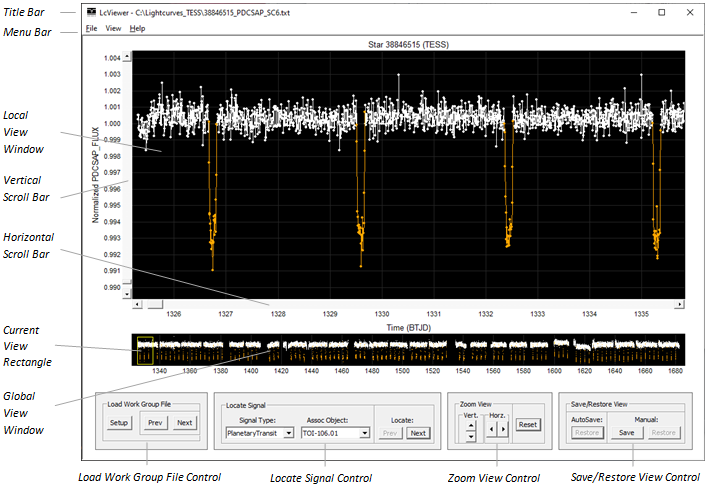}
\centering
\caption{The LcViewer Window}
\label{fig:LcViewer}
\end{figure*}

\begin{itemize}
\item The Local View Window showing the working area of the lightcurve. Time is displayed along the x-axis and normalized flux along the y-axis.
\item The Global View Window showing the full lightcurve.
\item The Current View Rectangle showing where the local view is located relative to the full lightcurve.
\item The Horizontal Scroll Bar for panning the lightcurve left and right.
\item The Vertical Scroll Bar for panning the lightcurve up and down.
\item The Load Work Group Control for setting up and loading lightcurve files from a work group. See section \ref{subsec:workgroup}.
\item The Locate Signal Control for finding and displaying signals in the lightcurve.
\item The Zoom View Control for zooming in or out of the lightcurve. 
\item The Save/Restore View Control for saving and restoring a view or to quickly return to the previous view if it was accidentally changed.
\end{itemize}


\subsection{Lightcurve Display}

\subsubsection{Optimizing the Initial View}

When a lightcurve is initially displayed, the view is scaled and shifted vertically such that high data point outliers are clipped. This optimization feature prevents excessive vertical shrinkage due to extreme outliers.

\subsubsection{Scaling of Data Points}

The lightcurve is normally drawn using white medium size data points connected with lines. However, if the data points are too close together, the application employs a congestion mitigation strategy whereby the data points are drawn in a smaller size and not connected with lines. This behavior can be be overridden using the Shift+z key which forces medium size points and connected lines.

\subsubsection{Rendering of Signals} \label{subsubsection:renderingsigs}

Imported signals from a signal library are highlighted in color. To help differentiate signals visually, each one is assigned a different color when the lightcurve file is loaded. All the instances of a periodic signal are drawn in the same color.

Signals are drawn by level number. All level 1 signals are drawn first followed by all level 2 signals. This ensures that level 2 signals are on top and visible.

A level 1 signal is a primary signal such as an planetary transit. A level 2 signal is a secondary signal of the primary signal such as an exomoon transit. The level number is set when the signal is created. See section \ref{subsubsec:createsig} for details.

\begin{figure}[ht]
\includegraphics[scale=0.53]{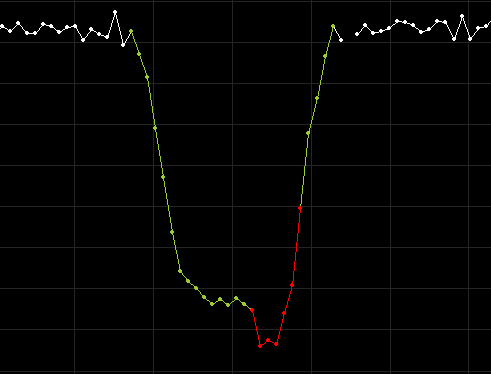}
\centering
\caption{Two overlapping planetary transit signals. The red area indicates the intersecting region.}
\label{fig:OverlappingSigs}
\end{figure}

If two signals with the same level number overlap, the intersecting region will be highlighted in red. See Figure \ref{fig:OverlappingSigs} for an example.

A partially or fully buried signal can be made fully visible by pressing the Ctrl+r key.

\subsubsection{Displaying Signal Markers}

\begin{figure}[ht]
\includegraphics[scale=0.7]{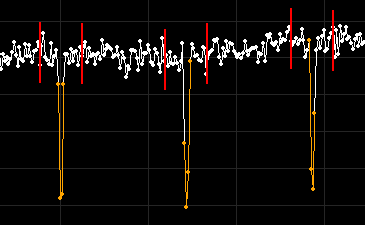}
\centering
\caption{Reference markers in red drawn around each instance of a signal using a time span of 7 signal durations.}
\label{fig:SignalMarkers}
\end{figure}

Left and right vertical markers can be displayed in red around each instance of the current signal in the lightcurve for visual reference purposes. See Figure \ref{fig:SignalMarkers} for an example. The time span between the left and right markers can specified in terms of signal durations, hours, days, periods, and Hill time scale.

\subsubsection{Displaying Background Flux} \label{subsubsec:backgrndflux}

\begin{figure}[ht]
\includegraphics[scale=0.53]{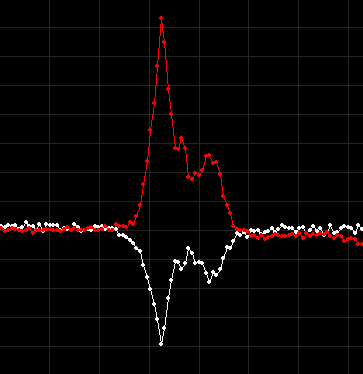}
\centering
\caption{Background flux shown in red. The inverted red signal above the white candidate signal indicates that the white signal was caused by a light-scattering event.}
\label{fig:BackgroundFlux}
\end{figure}

Background flux can be displayed using the Shift+b key to help identify light-scattering events in the lightcurve. See Figure \ref{fig:BackgroundFlux} for an example. Currently, this feature is only available for TESS and TASOC lightcurves.

\subsection{Lightcurve Navigation} \label{subsec:lcnav}

\subsubsection{Selecting a View}

A view may be selected in the Local or Global View Window by selecting a rectangular section of the lightcurve using the M1 button.

\subsubsection{Panning a Lightcurve}

A lightcurve can be panned in three ways: 1) By using the horizontal and vertical scroll bars, 2) by grabbing the local view with the M2 button and then dragging the view in the desired direction, and 3) by grabbing the Current View Rectangle with the M2 button and then dragging the rectangle in the desired direction.

\subsubsection{Auto-Scrolling a Lightcurve} \label{subsubsec:autoscroll}

Auto-scrolling a lightcurve is a way to automatically pan the lightcurve in association with an operation that requires a starting and ending location be selected with the mouse. If the ending location for the operation falls outside the current view, the view can be automatically scrolled by moving the cursor outside the Local View Window while the main operation key is held down. This is equivalent to panning the lightcurve with a scroll bar. The scroll speed is governed by the distance between the window boundary and the cursor. The greater the distance, the faster the scroll speed.

\subsubsection{Zooming In or Out of a Lightcurve}

A lightcurve can be zoomed in or out in three main ways: 1) By using the four horizontal and vertical zoom buttons in the Zoom View Control, 2) by using the four arrow keys on the keyboard, and 3) by pressing the Ctrl+M2 button and then moving the cursor in the direction to zoom. Of the three methods, the last one is the fastest and most versatile providing 360 degree zoom capability in one operation.

The ``Reset" button can be clicked to reset the view to the optimized initial view. Typically, this is the full lightcurve.

As yet another way to zoom, the M1 button can be double-clicked inside the Local or Global View Window to zoom in on a selected location or signal in the lightcurve. 

\subsubsection{Navigating the Instances of a Signal} \label{subsec:signav}

The instances of a defined signal in a lightcurve can be navigated sequentially using the Locate Signal Control. Each time the ``Next" or ``Prev" button is clicked, the next or previous instance of the selected signal is centered in the Local View Window and zoomed in. The horizontal and vertical scaling factor is preserved between instances.


\subsection{Real-Time Tracking}

\begin{figure}[ht]
\includegraphics[scale=0.8]{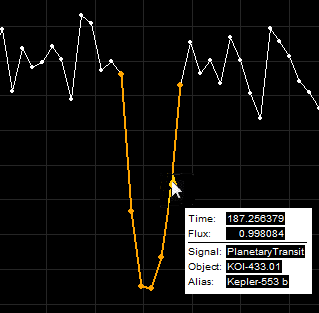}
\centering
\caption{The Tracking Information Box showing the time, flux, and signal at the cursor.}
\label{fig:TrackingInfoBox}
\end{figure}

The time, flux, and signal at the cursor can be monitored in real-time via the Tracking Information Box. See Figure \ref{fig:TrackingInfoBox} for an example.

If the cursor is positioned over a data point, the data point will be enlarged to indicate that the time and flux values have been snapped to the actual values in the lightcurve within six decimal places.

\subsection{Measuring the Interval Between Two Locations}

The time and flux interval between any two locations in the Local View Window can be measured using the Shift+M1 key. 

Pressing the key and then moving the mouse draws a measurement line between the starting and ending location. If the ending location lies outside the Local View Window, the lightcurve will be automatically scrolled as described in section \ref{subsubsec:autoscroll}.

\begin{figure}[ht]
\includegraphics[scale=0.8]{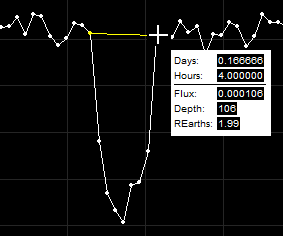}
\centering
\caption{A time based measurement. Here, the duration of a candidate signal is being measured.}
\label{fig:MeasureTime}
\end{figure}

\begin{figure}[ht]
\includegraphics[scale=0.8]{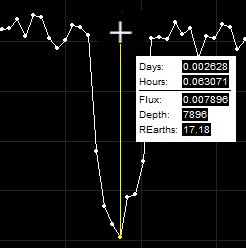}
\centering
\caption{A flux based measurement. Here, the depth and size of a candidate signal is being measured.}
\label{fig:MeasureFlux}
\end{figure}

The line is accompanied by a Measurement Information Box showing the time and flux interval between locations. The time interval is given in days and hours. The flux interval is given in flux units, depth (ppm), and REarth units. An example of a time based measurement is shown in Figure \ref{fig:MeasureTime} and a flux based measurement in Figure \ref{fig:MeasureFlux}.

\subsection{Recording Signals}

A signal may be recorded for any dip, peak, artifact or anomaly found in a lightcurve. There are no restrictions as what a signal can represent other than it must encompass at least three data points in the lightcurve including the end points. Signals may be periodic or non-periodic.

\newpage
\subsubsection{Creating a Signal} \label{subsubsec:createsig}

There are three ways to create a signal in LcViewer: 1) By enclosing the relevant data points with a bounding box using the Ctrl+M1 button, 2) by clicking the ``Create" button in the ``Find Periodic Signals" dialog box (see Figure \ref{fig:FindPeriodicSigsDlg}), and 3) by clicking the ``Create Signal" button in the ``Measure Candidate Signal" dialog box (see Figures \ref{fig:MeasureCandidateSigDlg2}, \ref{fig:MeasureCandidateSigDlg3}, and \ref{fig:MeasureCandidateSigDlg4}).

The ``Create Signal" dialog box will then open as shown in Figure \ref{fig:CreateSigDlg}. The fields may or may not be filled in with default values depending on the method used to start the process. At minimum, default values will be populated in the group boxes for Output Signal Library and Signal Times. The following fields can be set:

\begin{figure}[ht]
\includegraphics[scale=0.7]{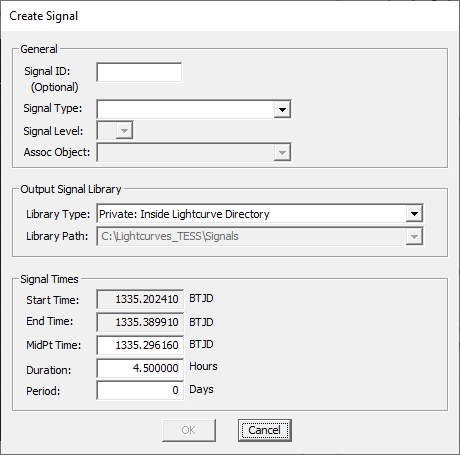}
\centering
\caption{The ``Create Signal" dialog box.}
\label{fig:CreateSigDlg}
\end{figure}

\begin{itemize}
\item Signal ID: An optional ID for the signal. Normally, this field is left blank.
\item Signal Type: The type of signal to create. A value can be selected from the drop-down list for an existing type or entered into the field for a new type. At minimum, the drop-down list will contain types ``PlanetaryTransit" and ``EB".
\item Signal Level: The level number for the signal \textbf{--} 1 for a primary signal such as an exoplanet and 2 for a secondary signal such as an exomoon.
\item Assoc Object: The ID of an object that is associated with the signal. A value can be selected from the drop-down list for an existing ID or entered into the field for a new ID. At minimum, the drop-down list will contain two items \textbf{--} the host star ID and a unique object ID of the form ``\textit{StarID.ObjectNum}". For example, 38846515.01.

An associated object is regarded as the source or cause for the signal based on the signal type. For example, if the signal type is “PlanetaryTransit”, the associated object would be the ID of the host planet.
\item Library Type: The type of signal library in which to store the signal. Options include ``Private: Inside Lightcurve Directory", ``Private: Outside Lightcurve Directory", and ``Public". Normally, the first option is used.
\item Library Path: The full path name of the target signal library.
\item Period: If the signal is periodic, the time interval in days between each instance of the signal. If the signal is non-periodic, a value of 0.
\end{itemize}

Non-periodic signals can be created with multiple instances by setting the signal type and associated object to the same value in each instance. LcViewer then groups all the instances together under one parent signal.

Once the dialog box is filled in and submitted, the signal will be added to the target signal library, instantiated in the lightcurve, and highlighted.

\subsubsection{Editing a Signal}

A public or private signal may be edited by moving the cursor over the signal and then pressing the Ctrl+e key. A bounding box for the signal will be drawn enabling the left and right sides to be adjusted with the mouse. After the sides are adjusted, a dialog box similar to Figure \ref{fig:CreateSigDlg} will be opened enabling the field values to be edited.

\subsubsection{Deleting a Signal} \label{subsubsec:deletesig}

A defined signal or instance thereof may be deleted by moving the cursor over the signal and pressing the Ctrl+d key.

\subsubsection{Moving a Signal} \label{subsubsec:movesig}

A defined signal or instance thereof may be moved by positioning the cursor over the signal, pressing the m+M2 button, and then dragging the bounding box for the signal to the target location using the mouse.


\subsection{Recording TTVs}

A TTV can be recorded for an instance of a defined periodic signal by moving, creating, or deleting the instance in the lightcurve as described in sections \ref{subsubsec:movesig}, \ref{subsubsec:createsig}, and \ref{subsubsec:deletesig} respectively. LcViewer interprets a move, create, or delete operation for an instance of a periodic signal as a TTV alignment request and records the information in a TTV library rather than in the parent signal library.


\subsection{Querying the Properties of a Defined Signal}

\begin{figure}[ht]
\includegraphics[scale=0.72]{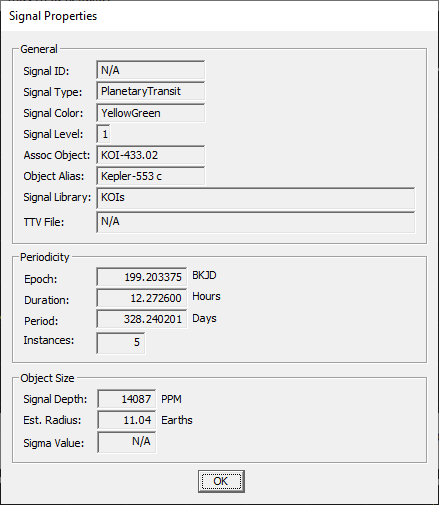}
\centering
\caption{The ``Signal Properties" dialog box.}
\label{fig:SignalPropsDlg}
\end{figure}

The properties of a defined signal in a lightcurve can be queried by moving the cursor over the signal and pressing the Ctrl+q key. The ``Signal Properties" dialog box will then open as shown in Figure \ref{fig:SignalPropsDlg}.

\subsection{Editing a Lightcurve}

A section of a lightcurve can be removed by marking the target section with a rectangle using the Alt+M2 button. All data points inside the rectangle will be deleted from the lightcurve. This feature is commonly used to remove low quality sections of a lightcurve prior to manual detrending (see section \ref{subsec:detrend}). 

The edited lightcurve can optionally be saved to the same lightcurve file or a different one.


\subsection{Manually Detrending a Lightcurve} \label{subsec:detrend}

A lightcurve can be manually detrended to reduce large-scale fluctuations so that signals are easier to detect. 

The detrending process is started by pressing the Shift+d key or by selecting the equivalent command from the menu bar. An initial green trend line will be displayed in the lightcurve accompanied by the ``Detrend Lightcurve" dialog box for controlling the operation. See Figure \ref{fig:DetrendLcDlg1} for an example. The dialog box can be moved outside the LcViewer window so that it does not interfere with the lightcurve. The position will be remembered the next time the dialog box is opened.

At this point the user may perform any of the operations below:
\begin{itemize}
\item Adjust the checkboxes for removing high and low single point outliers from the lightcurve prior to detrending. If the first box is checked, the worst 1\% of high outliers are removed. If the second box is checked, the worst 0.05\% of low outliers are removed. Normally, both boxes are checked.
\item Select a trend line fitting method \textbf{--} Moving Median or Spline. Moving Median is the default. Spline is not recommended for time spans over 100 days.
\item Adjust the trend line fitting level. Levels range from 1 to 25. The higher the value, the tighter the fit and the flatter the lightcurve will be after detrending. The default value is 16.

While adjusting the level, the green trend line can be visually inspected for underfitting and overfitting. Overfitting may severely attenuate or destroy signals of interest when the lightcurve is detrended. See Figure \ref{fig:OverFitting} for an example. 

\begin{figure}[ht]
\includegraphics[scale=0.72]{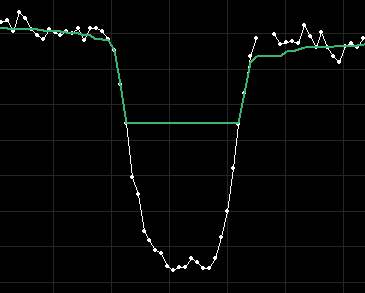}
\centering
\caption{Example of overfitting. The green trend line extends well into the white signal which will cause the signal to be severely attenuated when the lightcurve is detrended.}
\label{fig:OverFitting}
\end{figure}

\item Click the ``Detrend" button to detrend the lightcurve based on the fitted green line. See Figure \ref{fig:DetrendLcDlg2} for an example.

\item Click the ``Redo" button to reset the lightcurve to its previous state so that the detrending operation can be retried with different settings.
\item Click the ``OK" button to accept the detrended results.
\item Optionally save the detrended lightcurve to the same lightcurve file or a different one. 
\end{itemize}


\subsection{Vetting Signals Found by LcSignalFinder} \label{subsec:vettingsigs}

If periodic signals were found previously by LcSignalFinder, the signals must be vetted and then recorded in a signal library before they can be used.

To start the process, a work group must first be set up for the lightcurves files having periodic signals from LcSignalFinder. This can be done by clicking the ``From Signals Found" button in the ``Setup Work Group" dialog box as shown in Figure \ref{fig:SetupWorkGroupDlg}.

Each file from the work group can then be opened sequentially by clicking the ``Next" button at the bottom left corner of the LcViewer window or by pressing the Shift+n key. 

\begin{figure*}[htb!]
\includegraphics[scale=0.70]{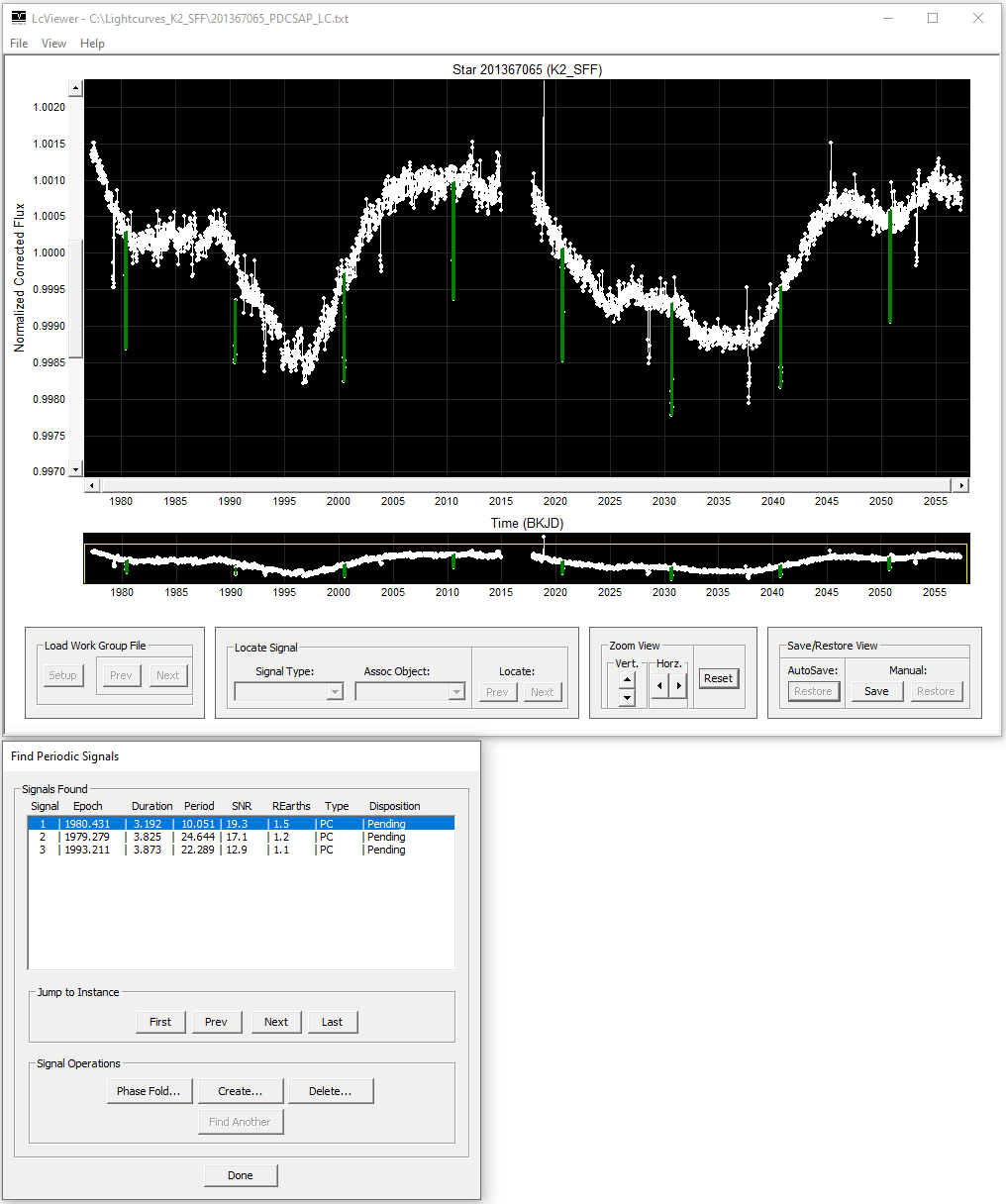}
\caption{A lightcurve showing a periodic signal to be vetted in green accompanied by the ``Find Periodic Signals" dialog box for controlling the operation.}
\label{fig:FindPeriodicSigsDlg}
\end{figure*}

Upon opening a file, LcViewer will display the ``Find Periodic Signals" dialog box for controlling the vetting operation. See Figure \ref{fig:FindPeriodicSigsDlg} for an example.  The dialog box can be moved outside the LcViewer window so that it does not interfere with the lightcurve. The position will be remembered the next time the dialog box is opened.

Each signal in the dialog box will show the reference epoch, signal duration in hours, period in days, signal-to-noise ratio (SNR), signal size in REarth units, type of signal, and current disposition.

At this point the user can perform the following operations on each signal in the dialog box:
\begin{itemize}
\item Select a working signal from the list. Each instance of the signal will be marked with a green rectangle in the lightcurve for identification purposes.
\item Using the four navigation buttons in the dialog box, iterate through each instance of the working signal to visually check the alignment between the green rectangle and the actual instance.
\item If the instances are misaligned, use the m+M2 button to manually align the first or last instance in the lightcurve whichever one has the greatest deviation. This will automatically adjust the reference epoch and period resulting in a better overall fit.
\item Click the ``Phase Fold" button to phase fold the periodic signal and adjust the signal duration. See section \ref{subsec:phasefolding} for more information.
\item If the signal is viable, click the ``Create" button to record the signal in a library. The ``Create Signal" dialog box will then open with all the required fields filled in (see Figure \ref{fig:CreateSigDlg}). Upon clicking the ``OK" button, the signal will be added to the target signal library, instantiated in the lightcurve, and highlighted.

If the signal is not viable, click the ``Delete" button to delete the candidate signal from the dialog box and lightcurve.
\end{itemize}

\subsection{Detecting and Recording Periodic Signals}

There are three methods in LcViewer for detecting and recording periodic signals in a lightcurve \textbf{--} automatic, semi-automatic, and manual. Each method is described below.

\subsubsection{Method 1: Automatic Signal Detection}

In method 1, the current lightcurve is searched for periodic signals using BLS in a manner similar to LcSignalFinder. 

\begin{figure}[htb!]
\includegraphics[scale=0.72]{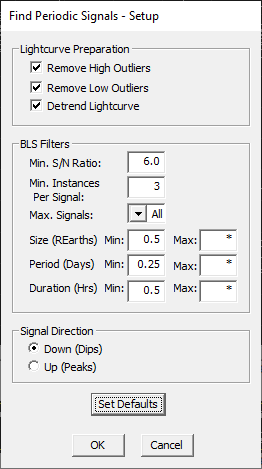}
\centering
\caption{The ``Find Periodic Signals - Setup" dialog box.}
\label{fig:FindPeriodicSigsSetupDlg}
\end{figure}

The process is started by pressing the Ctrl+s key. The ``Find Periodic Signals - Setup" dialog box will then open as shown in Figure \ref{fig:FindPeriodicSigsSetupDlg}. The settings in the dialog box are identical to the ones for LcSignalFinder. See section \ref{subsubsec:findermainsetup} for details.

LcViewer will then search for the first periodic signal in the lightcurve using a process similar to that described in steps 5-10 of section \ref{subsubsec:finderprocess}. If a signal is found, the ``Find Periodic Signals" dialog box will be opened and the instances of the signal marked with green rectangles like that shown in Figure \ref{fig:FindPeriodicSigsDlg}.

The user can then examine each instance, adjust the reference epoch and period, phase fold the signal to adjust the signal duration, and then record the signal in a library (if viable) as described in section \ref{subsec:vettingsigs}. The ``Find Another" button can then be clicked to search for the next periodic signal in the lightcurve. The process is repeated until all the periodic signals have been found and dispositioned.

\subsubsection{Method 2: Semi-Automatic Signal Detection} \label{subsubsec:semiautodetect}

In method 2, the user identifies a candidate reference signal in the lightcurve. Using the reference signal as a template, LcViewer then searches the lightcurve for all matching signals that are periodic with respect to the reference signal and returns the periods found.

The process is started by enclosing the reference signal with a rectangle using the Alt+M1 button. The ``Measure Candidate Signal" dialog box will then open as shown in Figure \ref{fig:MeasureCandidateSigDlg1}. The dialog box can be moved outside the LcViewer window so that it does not interfere with the lightcurve. The position will be remembered the next time the dialog box is opened.

Upon clicking the ``Check Periodicity" button in the dialog box, LcViewer scans the lightcurve for all signals that closely match the reference signal in terms of signal duration and depth. Next, it examines the signals found to determine whether any are periodic with respect to the reference signal. If found, the candidate periods are listed in the dialog box with the strongest periods at the top based on the matching signal count and percentage. See Figure \ref{fig:MeasureCandidateSigDlg2} for an example.

At this point the user can perform the following operations on the candidate periods:

\begin{itemize}
\item Select a working period from the list. Each instance of the signal at the selected period will then be marked with a rectangle in the lightcurve \textbf{--} green if the instance matches the reference signal and red if not.
\item Using the five navigation buttons in the dialog box, iterate through each instance of the working signal to visually check the alignment between the rectangle and the actual instance.
\item If the instances are misaligned, use the m+M2 button to manually align the first or last instance in the lightcurve whichever one has the greatest deviation. This will automatically adjust the reference epoch and period resulting in a better overall fit.
\item Click the ``Phase Fold" button to phase fold the working signal. See section \ref{subsec:phasefolding} for more information.
\item If the periodic signal is viable, click the ``Create" button. The ``Create Signal" dialog box will then open with all the required fields filled in (see Figure \ref{fig:CreateSigDlg}). Upon clicking the ``OK" button, the signal will be added to the target signal library, instantiated in the lightcurve, and highlighted.
\end{itemize}


\subsubsection{Method 3: Manual Signal Detection}

In method 3, the user identifies two adjacent reference signals in the lightcurve. Based on the time interval between reference signals, LcViewer calculates a period and then instantiates the signal across the lightcurve for the period.

The process is started by enclosing the first reference signal with a rectangle using the Alt+M1 button. The best results are achieved if the reference signal resides near the middle of the lightcurve. The ``Measure Candidate Signal" dialog box will then open as shown in Figure \ref{fig:MeasureCandidateSigDlg1}.

The second reference signal is then enclosed with a rectangle using Alt+M1. The second signal must be adjacent to the first signal, either before it or after it with no intervening gaps. LcViewer then determines the period for the signal by calculating the time interval between the two reference signals. Based on the period, LcViewer generates an entry in the list box, instantiates the signal in the lightcurve, and marks each instance of the signal with a yellow rectangle. See Figure \ref{fig:MeasureCandidateSigDlg3} for an example.

The user can then examine each instance, adjust the reference epoch and period, phase fold the signal, and then record the signal in a library (if viable) as described in section \ref{subsubsec:semiautodetect}.

\subsection{Phase Folding a Periodic Signal} \label{subsec:phasefolding}

A periodic signal may be phase folded to obtain a composite signal for study. 
There are three ways in which to start the operation: 1) By pressing the Ctrl+f key inside an instance of the signal, 2) by clicking the ``Phase Fold" button inside the ``Find Periodic Signals" dialog box (see Figure \ref{fig:FindPeriodicSigsDlg}), and 3) by clicking the ``Phase Fold Signal" button inside the ``Measure Candidate Signal" dialog box (see Figures \ref{fig:MeasureCandidateSigDlg2} and \ref{fig:MeasureCandidateSigDlg3}).

\subsubsection{The ``Phase Fold Setup" Dialog Box} \label{subsubsec:phasefoldsetup}

Upon starting the operation, the ``Phase Fold Setup" dialog box will be opened as shown in Figure \ref{fig:PhaseFoldSetupDlg}. Settings include:

\begin{figure}[ht]
\includegraphics[scale=0.8]{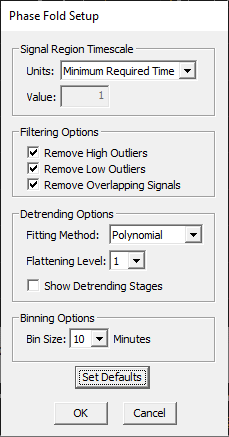}
\centering
\caption{The ``Phase Fold Setup" dialog box.}
\label{fig:PhaseFoldSetupDlg}
\end{figure}

\begin{itemize}
\item The signal region timescale indicating the time span to be phase folded for each instance of the periodic signal in the lightcurve centered at the midpoint of the instance.

The timescale is governed by two parameters  \textbf{--} time units and time value. Available time units include hours, days, periods, signal durations, Hill timescale, and minimum required time (the default). The time value indicates the number of units in the signal region. For example, 7 signal durations.
\item Filtering options for removing high data point outliers, low data point outliers, and overlapping signals from the lightcurve prior to detrending. Normally all three options are selected.
\item Detrending options for setting the trend line fitting method and flattening level.

Two fitting methods are available \textbf{--} Polynomial and Spline. The default is Polynomial for signal regions under 10 days and Spline for signal regions above 10 days.

Flattening levels range from 1 to 25. The higher the value, the tighter the fit and the flatter the signal region will be after detrending. The default is 1 (no flattening).
\item An option for showing the detrending stages on each instance of the signal to check for underfitting and overfitting. Normally, this option is not selected.
\item The bin size in minutes for binning the phase folded data points. Available sizes include 1, 2, 3, 4, 5, 6, 10, 12, 15, 20, and 30 minutes. The default is 10 minutes.
\end{itemize}

For a typical user, using the default values will provide high quality phase folded results. No adjustments are normally needed.

\subsubsection{The Phase Folding Process}

Once the dialog box is submitted, LcViewer will perform the following tasks subject to the settings:

\begin{enumerate}
\item Removes the worst 1\% of high outliers and the worst 0.05\% of low outliers from the lightcurve.
\item Removes the data points for all overlapping signals in the lightcurve to prevent the signals from appearing in the phase folded lightcurve.
\item For each instance of the target signal in the lightcurve, a) fits a trend line through the signal region excluding the instance itself which is masked out, b) detrends the signal region based on the fitted trend line,
c) normalizes the flux values in the signal region to a mean value of 1.0, d) checks the quality of the detrended result and rejects the instance if the quality is low, and e) adds the normalized data points from the signal region to a collection area.
\item Bins the data points from the collection area.
\item Displays the phase folded lightcurve and an accompanying dialog box for controlling the phase fold operation.
\end{enumerate}

\begin{figure*}[htb!]
\includegraphics[scale=0.70]{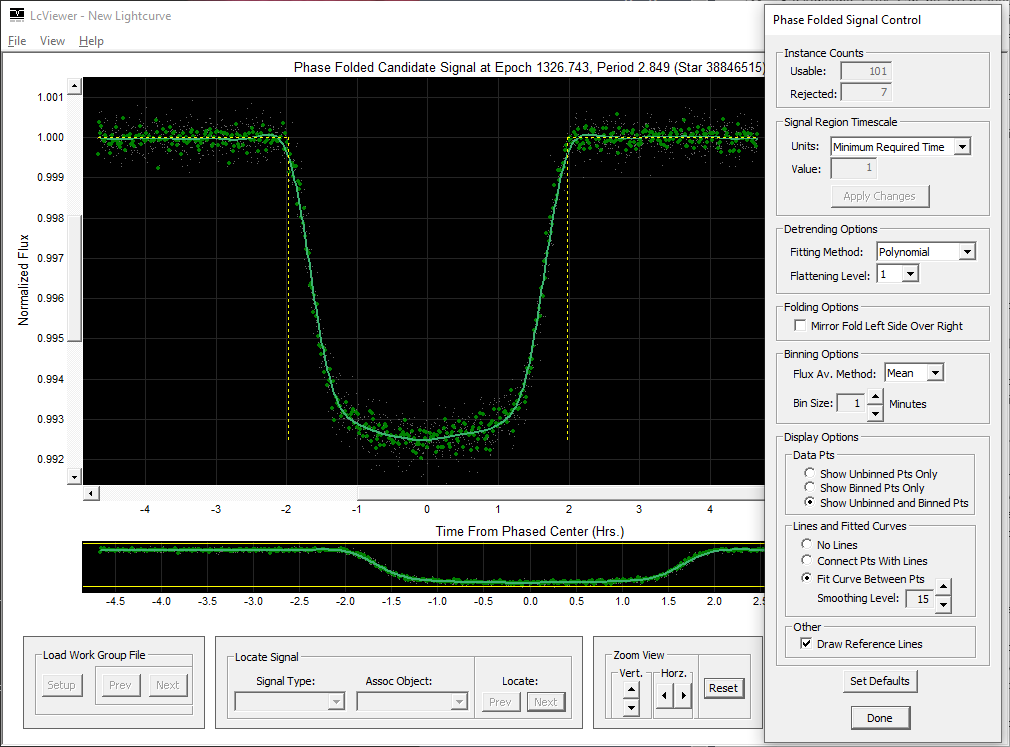}
\caption{A phase folded lightcurve and accompanying dialog box.}
\label{fig:PhaseFoldControlDlg}
\end{figure*}

\subsubsection{The Phase Folded Lightcurve}

A sample phase folded lightcurve is shown in Figure \ref{fig:PhaseFoldControlDlg}. The small grey dots represent the unbinned data points. The large green dots represent the binned data points. The green line shows the average fit through the binned points.

Time is displayed along the x-axis with the midpoint of the phase folded signal located at time = 0. Based on the signal region timescale, time units may be shown in hours, days, or periods. Normalized flux is displayed along the y-axis.

A horizontal reference line is drawn at a flux value of 1.0. The left and right vertical reference lines represent the extent of the phase folded signal. The time span between the lines is the signal duration.

\subsubsection{The ``Phase Folded Signal Control" Dialog Box} \label{subsubsection:phasefolddlg}

The accompanying dialog box shown in Figure \ref{fig:PhaseFoldControlDlg} can be used to fine-tune the original phase fold settings and to control how the lightcurve is displayed. The dialog box can be moved outside the LcViewer window so that it does not interfere with the lightcurve. The position will be remembered the next time the dialog box is opened.

The Signal Region Timescale, Fitting Method, Flattening Level, and Bin Size were all described in section \ref{subsubsec:phasefoldsetup} and so will not be repeated here. Additional settings include the following:

\begin{itemize}
\item An option for folding the left side data points over the right side to check for symmetry and to combine points.
\item The target method for averaging flux values when binning data points. Options include Mean and Median. The default is Mean.
\item The type of data points to display in the lightcurve. Options include ``Show Unbinned Pts Only", ``Show Binned Pts Only", and ``Show Unbinned and Binned Pts". The last option is the default.
\item The type of line or fitted curve to display through the binned data points. Options include ``No Lines", ``Connect Pts With Lines", and ``Fit Curve Between Pts". 

For the last option, a spline curve is used for signal regions below 10 days and a moving median curve is used for signal regions above 10 days. The Smoothing Level control adjusts the smoothing factor. Levels range from 1 to 33. The lower the value, the tighter the fit. The higher the value, the looser the fit. The default value is 17 which provides average smoothing.
\item An option for turning the reference lines on and off.
\end{itemize}

For a typical user, using the default values will provide high quality phase folded results. No adjustments are normally needed.

\subsubsection{Adjusting the Signal Duration}

\begin{figure}[ht]
\includegraphics[scale=0.8]{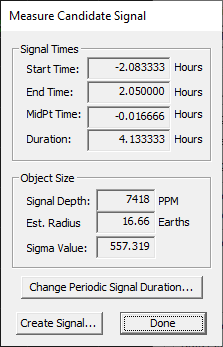}
\centering
\caption{The ``Measure Candidate Signals" dialog box for a phase folded lightcurve.}
\label{fig:MeasureCandidateSigDlg4}
\end{figure}

The duration of the phase folded signal can be adjusted by placing a bounding box around the signal using the Alt+M1 button. The ``Measure Candidate Signal" dialog box will open as shown in Figure \ref{fig:MeasureCandidateSigDlg4}.

Upon clicking the ``Change Periodic Signal Duration" button in the dialog box, LcViewer will change the signal duration based on the bounding box and then adjust the left and right vertical reference lines accordingly. When the ``Phase Folded Signal Control" dialog box is closed, the adjusted duration will be applied to the source signal that was phase folded.

\subsubsection{Auxiliary Operations} \label{subsubsection:phasefoldauxops}

Most operations that can be performed in a regular lightcurve can be done in a phase folded lightcurve. The user can 1) save and load phase folded lightcurve files, 2) navigate the lightcurve (select views, pan, and zoom), 3) measure intervals, 4) measure, create, edit, delete, and move signals, 5) query the properties of a defined signal, and 6) query the properties of the host star.


\subsection{Querying the Stellar Properties for the Host Star}

\begin{figure}[ht]
\includegraphics[scale=0.8]{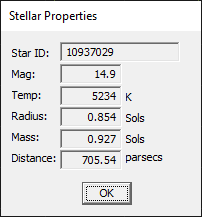}
\centering
\caption{The ``Stellar Properties" dialog box.}
\label{fig:StellarPropsDlg}
\end{figure}

The stellar properties for the host star can be queried from MAST or NEA by pressing the Shift+q key or by selecting the equivalent command from the menu bar. The ``Stellar Properties" dialog box will then open as shown in Figure \ref{fig:StellarPropsDlg}.

\subsection{Supporting Documentation}
A user guide and quick reference manual for LcViewer can be accessed via the Shift+g and Shift+m keys respectively or by selecting the equivalent commands from the menu bar. Both documents can also be accessed from the LcTools installation directory.

\newpage
\section{\large{L\lowercase{c}R\lowercase{eporter}}} \label{sec:reporter}

\subsection{Main Features}

LcReporter creates an Excel report for the signals recorded by LcViewer. The signals from multiple public and private signal libraries may be combined together into one report. Separate worksheets in the file are provided for regular signals and phase folded signals.

Data columns on the regular signals worksheet include star ID, signal ID, signal type, associated object, signal level, start time, end time, midpoint time, duration, period, and signal file. See Figure \ref{fig:Signals} for an example.

Data columns on the phase folded signals worksheet include all those for regular signals (excluding the period) plus data columns for the periodic signal that was phase folded. These include the source signal type, associated object, first epoch, duration, and period. 

The signals in a worksheet can be sorted and filtered to help organize and analyze the data collected.

\subsection{Supporting Documentation}

A user guide for LcReporter can be accessed from the LcTools installation directory.

\section{Summary} \label{sec:summary}

LcTools is a Windows based software system for finding and recording signals of interest in large sets of lightcurves for the TESS, K2, and Kepler projects in addition to the TASOC, K2SFF, and EVEREST High Level Science Products.

The system can be used to generate, view, edit, and detrend lightcurves and to detect, record, measure, locate, query, highlight, and phase fold signals. Signals can be recorded for any type of phenomena or artifact whether periodic or non-periodic. The system can also be used to record TTVs, measure time and flux intervals, query stellar properties, and generate signal reports.

The software is free and can be obtained from the lead author by request at \href{mailto:aschmitt@comcast.net}{aschmitt@comcast.net}. For additional information on the product, see the LcTools Product Description\footnote{\url{https://sites.google.com/a/lctools.net/lctools/lctools-product-description}}.


\appendix

\section{Run-Time Requirements} \label{sec:rtreqs}

The following hardware and software is required to run LcTools:
\begin{itemize}
\item Windows OS (XP, Vista, 7, 8, 10).
\item 2.7 GHz CPU.
\item 5 GB memory.
\item 100 GB free disk space.
\item 1024 x 768 screen resolution.
\item 2-Button mouse or equivalent.
\item High-speed Internet connection.
\item Microsoft Word.
\item Microsoft Excel (if using LcReporter).
\item Google Drive (if using public signal libraries or public TTV libraries).
\end{itemize}

\newpage
\acknowledgments

We wish to thank the following individuals for their assistance in testing the product and contributing ideas for new features: Steve Hutcheon, Tom Jacobs, Kian Jek, Rebekah and Jennifer Kahn, Martti Kristiansen, Daryll LaCourse, Mark Omohundro, Hans Schwengeler, Johann Sejpka, Alton Spencer, Arvin Tan, Christopher Tanner, and Ivan Terentev.

A special thanks to Scott Fleming for providing technical assistance in accessing data at MAST.

J.H. acknowledges support from NASA grants NNX17AB61G and 80NSSC19K0386. Initial development of the BLS module that has been incorporated into LcTools was supported by NASA grant NNX14AE87G.

\bibliography{LcTools}{}
\bibliographystyle{aasjournal}


\begin{figure*}[ht]
\includegraphics[scale=0.70]{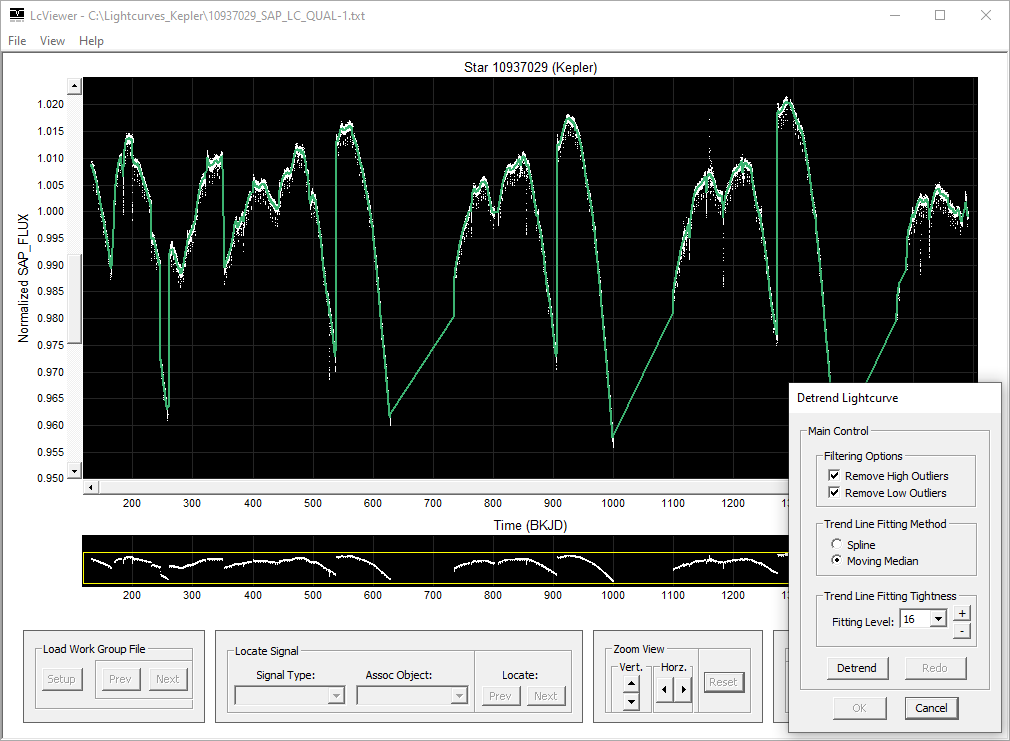}
\caption{A lightcurve to be detrended accompanied by the ``Detrend Lightcurve" dialog box. The green line is the fitted trend line. See Figure \ref{fig:DetrendLcDlg2} for the detrended result.}
\label{fig:DetrendLcDlg1}
\end{figure*}

\begin{figure*}[ht]
\includegraphics[scale=0.70]{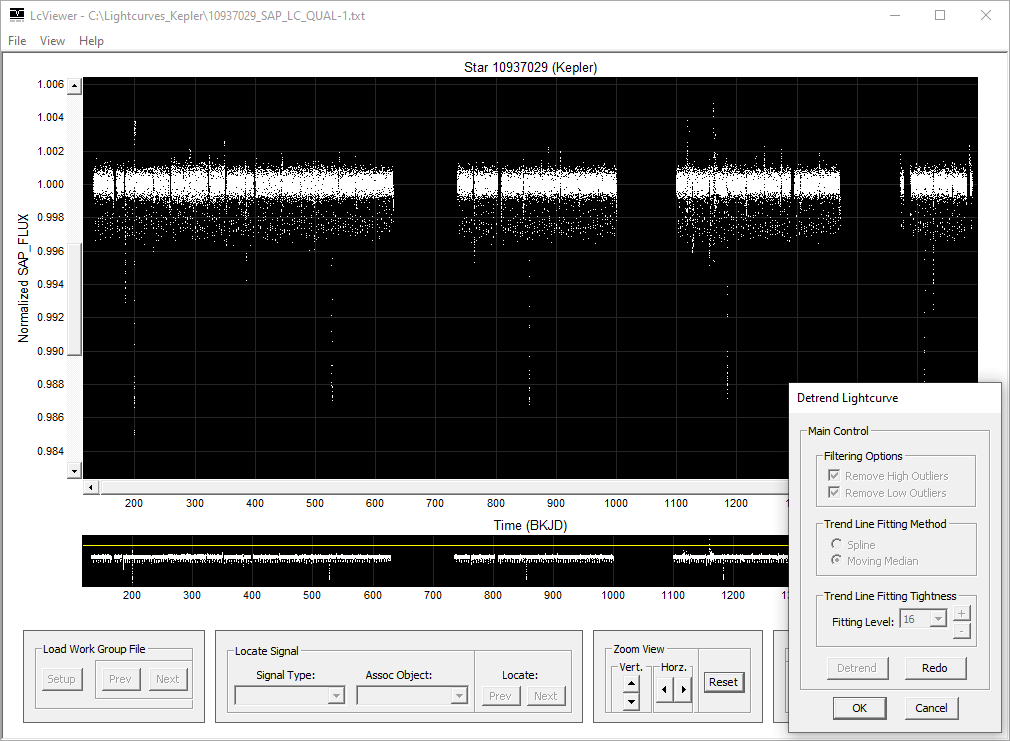}
\caption{The detrended result for the original lightcurve shown in Figure \ref{fig:DetrendLcDlg1}.}
\label{fig:DetrendLcDlg2}
\end{figure*}

\begin{figure*}[ht]
\includegraphics[scale=0.70]{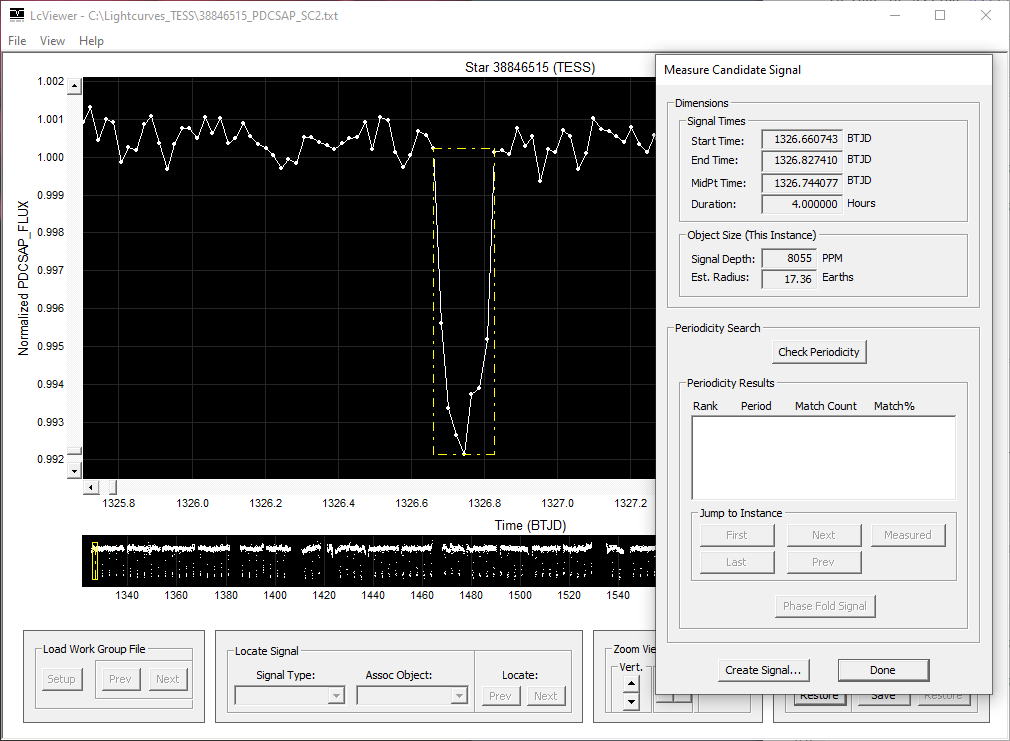}
\caption{A marked reference signal in yellow accompanied by the ``Measure Candidate Signal" dialog box. Applicable to signal detection methods 2 and 3.}
\label{fig:MeasureCandidateSigDlg1}
\end{figure*}

\begin{figure*}[ht]
\includegraphics[scale=0.70]{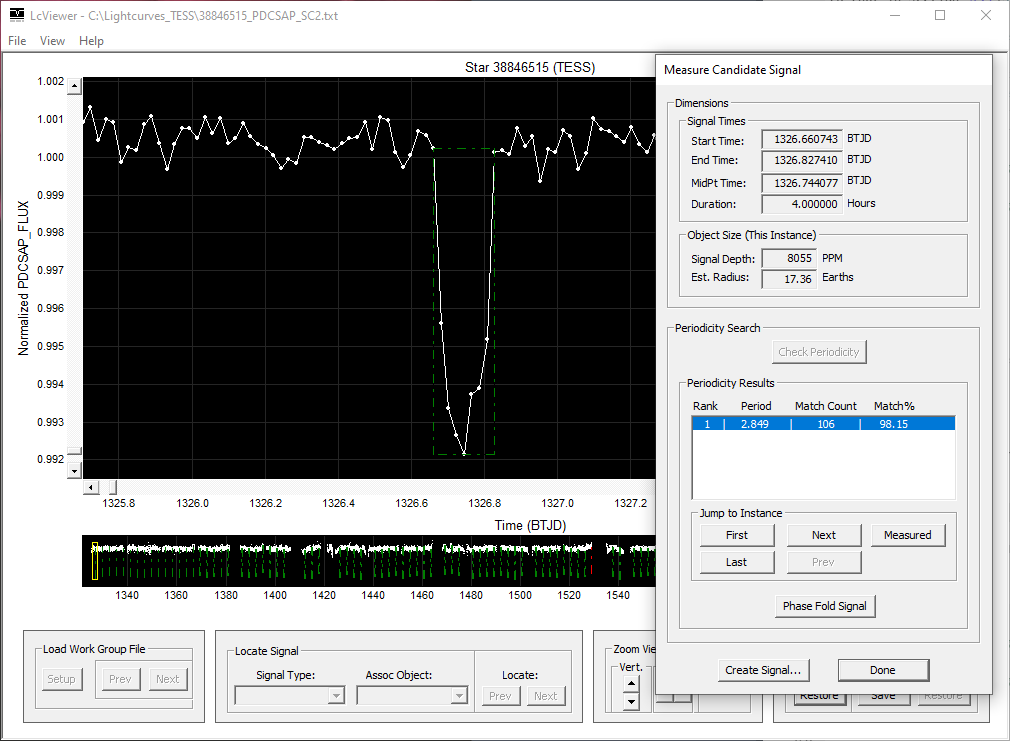}
\caption{The ``Measure Candidate Signal" dialog box showing one detected period for the reference signal. Each instance of the periodic signal in the lightcurve is marked with a rectangle \textbf{--} green if it matches the reference signal and red if not. Applicable to signal detection method 2 only.}
\label{fig:MeasureCandidateSigDlg2}
\end{figure*}

\begin{figure*}[ht]
\includegraphics[scale=0.70]{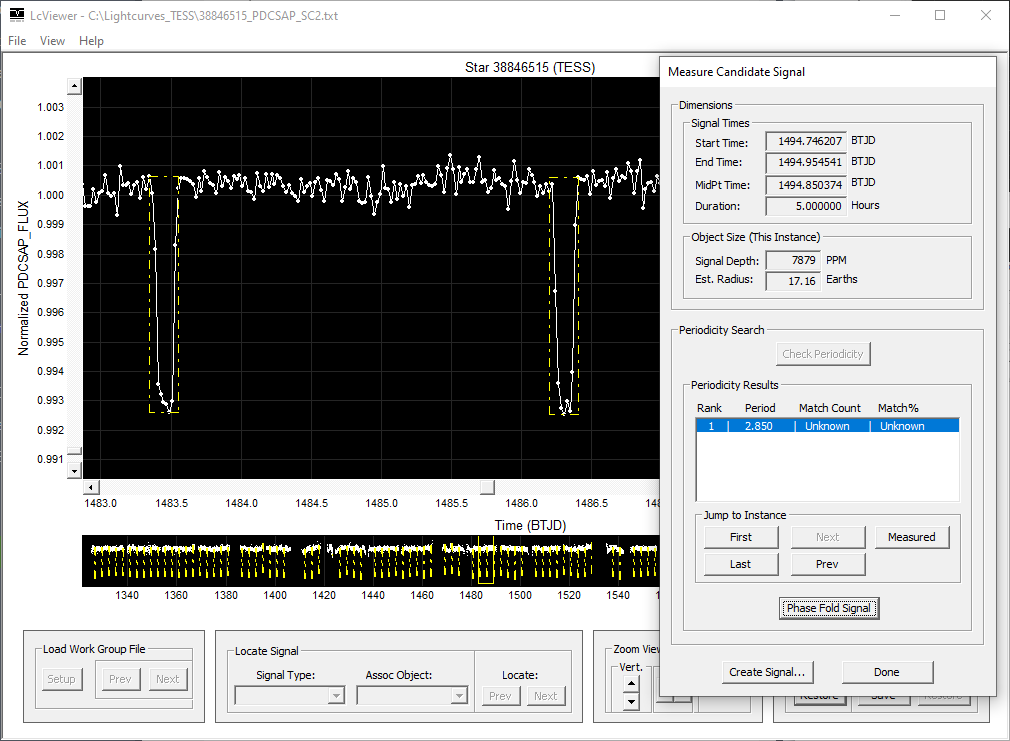}
\caption{The ``Measure Candidate Signal" dialog box showing the period for the two adjacent reference signals. Each instance of the periodic signal in the lightcurve is marked with a yellow rectangle. Applicable to signal detection method 3 only.}
\label{fig:MeasureCandidateSigDlg3}
\end{figure*}

\begin{figure*}[ht]
\includegraphics[scale=0.53]{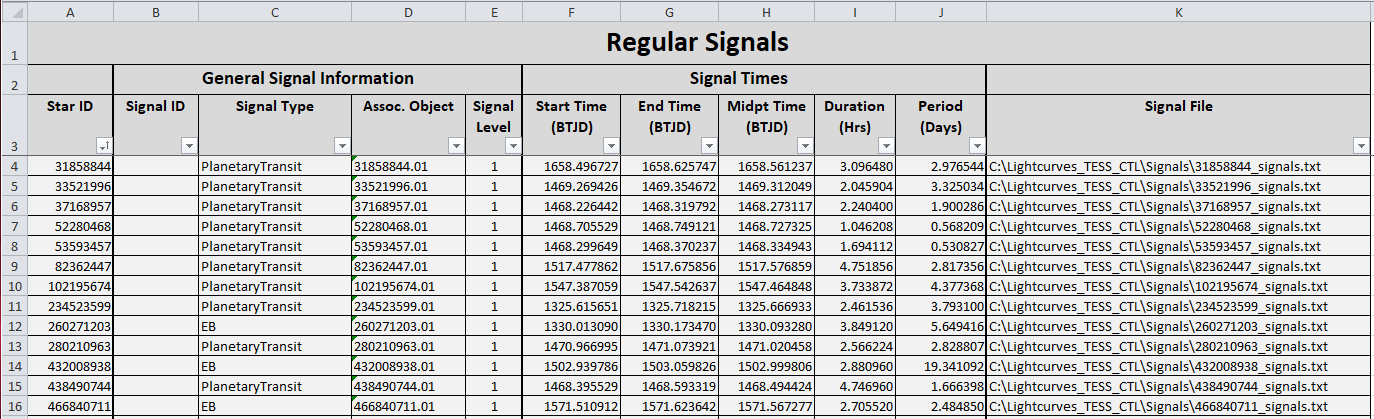}
\centering
\caption{The Regular Signals worksheet produced by LcReporter.}
\label{fig:Signals}
\end{figure*}

\begin{figure}[ht]
\includegraphics[scale=0.75]{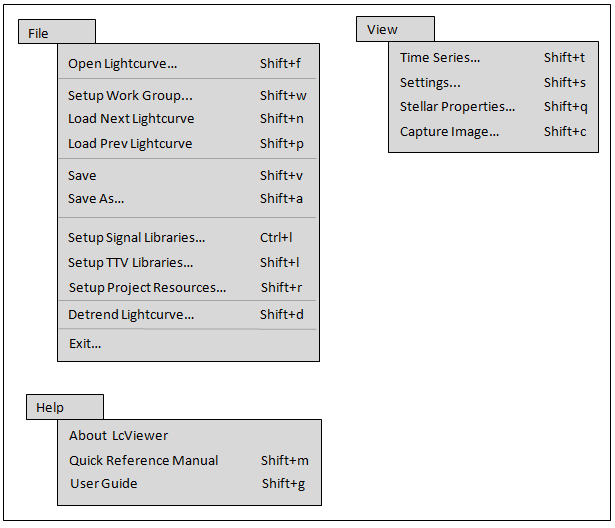}
\centering
\caption{Menu bar commands in LcViewer}
\label{fig:MenuBar}
\end{figure}

\begin{figure}[ht]
\includegraphics[scale=0.85]{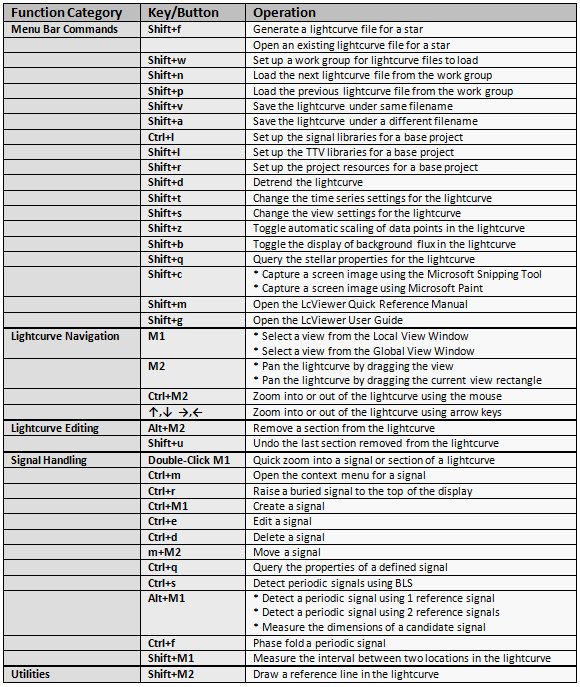}
\centering
\caption{Hot keys and hot buttons in LcViewer}
\label{fig:Hotkeys}\end{figure}


\end{document}